\documentclass[a4paper,twoside]{article}

\usepackage{multirow}
\usepackage{makecell}
\usepackage{url}
\usepackage{tikz}
\usepackage{stfloats}

\usepackage{epsfig}
\usepackage{subcaption}
\usepackage{calc}
\usepackage{amssymb}
\usepackage{amstext}
\usepackage{amsmath}
\usepackage{amsthm}
\usepackage{multicol}
\usepackage{pslatex}
\usepackage{apalike}
\usepackage{algorithm2e}
\usepackage[bottom]{footmisc}
\usepackage{SCITEPRESS}     

\begin{document}

\title{Stance Entropy and Judge Evaluation in Multi-Agent Debates}


\author{\authorname{Anonymous ICAART Submission}
\affiliation{}
\email{}
}

\author{\authorname{Onur Bilgin, Abdullah As Sami and John Licato}
\affiliation{Advancing Machine and Human Reasoning (AMHR) Lab\\
Bellini College of Artificial Intelligence, Cybersecurity and Computing\\University of South Florida}
\email{\{onurbilgin, sami20, licato\}@usf.edu}
}



\abstract{
    With the expanding adoption of multi-agent systems for complex reasoning and decision-making tasks, identifying reliable, accurate, and efficient methods for resolving debate outcomes has become increasingly important. We evaluate different decision-making strategies, including majority voting, consensus, and verbal confidence. In addition, we introduce a judge-based approach and an entropy-based decision protocol, derived from Shannon entropy, where debates are terminated when the entropy of the agents' stance distribution falls below a predefined entropy threshold, indicating convergence.
    Results demonstrate that the entropy-based method can serve as an efficient alternative to majority voting, especially in multiple-choice tasks with a larger number of agents, while employing language model-backed judges can improve accuracy when used strategically. Overall, we present a set of decision strategies that adapt to the interplay between dataset characteristics, agent count, and efficiency requirements.
}

\onecolumn \maketitle \normalsize \setcounter{footnote}{0} \vfill

\section{\uppercase{Introduction}}
\label{sec:introduction}

    Recent advances in large language models (LLMs) have drawn growing interest in multi-agent debate (MAD) systems, where multiple agents reason and argue collaboratively or from opposing perspectives towards a solution. MAD systems have improved the accuracy and justifiability of decision-making while promoting viewpoint diversity and enabling the exchange of arguments between opposing viewpoints \cite{du2024improving,liang-etal-2024-encouraging}.
    However, MAD systems cause increased computational overhead, making a predetermined decision protocol essential, which should be triggered once the predefined objective is reached, to save computational resources.
    This leads to the following challenge: when should a multi-agent debate be terminated, how can we define an appropriate decision protocol, and how should we decide the outcome of the debate?

    Terminating a debate too early risks preventing the agents from applying deeper reasoning or leaving the opposing arguments unresolved. Continuing it longer may lead to unnecessary computational costs. The state-of-the-art MAD systems rely on a fixed number of iterations or turns with predetermined early stopping rules, such as majority vote or a judge to evaluate the debate \cite{choi2025debate,kaesberg2025votingconsensus}. However, each decision protocol favors certain debate configurations, which can be counterproductive. For example, a majority vote may be less effective with an even number of agents due to the risk of ties, while a unanimous consensus rule may be unsuitable in applications with a large number of agents due to the small likelihood of reaching unanimity. The decision protocol should be chosen according to the debate setting.
    Another key consideration is the scalability of the protocols with increasing numbers of agents. Diverse personas, beliefs, and skills allow more perspectives to actively contribute to the debate \cite{zheng2024helpful,bilgin2026effect,jiang2026sok}. Decision protocols should be able to identify the most appropriate answer among the collective perspectives.
    An entropy-based protocol that aggregates the distribution of agents' decisions may be more effective as the number of agents increases.


    \begin{figure}[htb!] 
              \centering
                \includegraphics[width=1.0\linewidth]{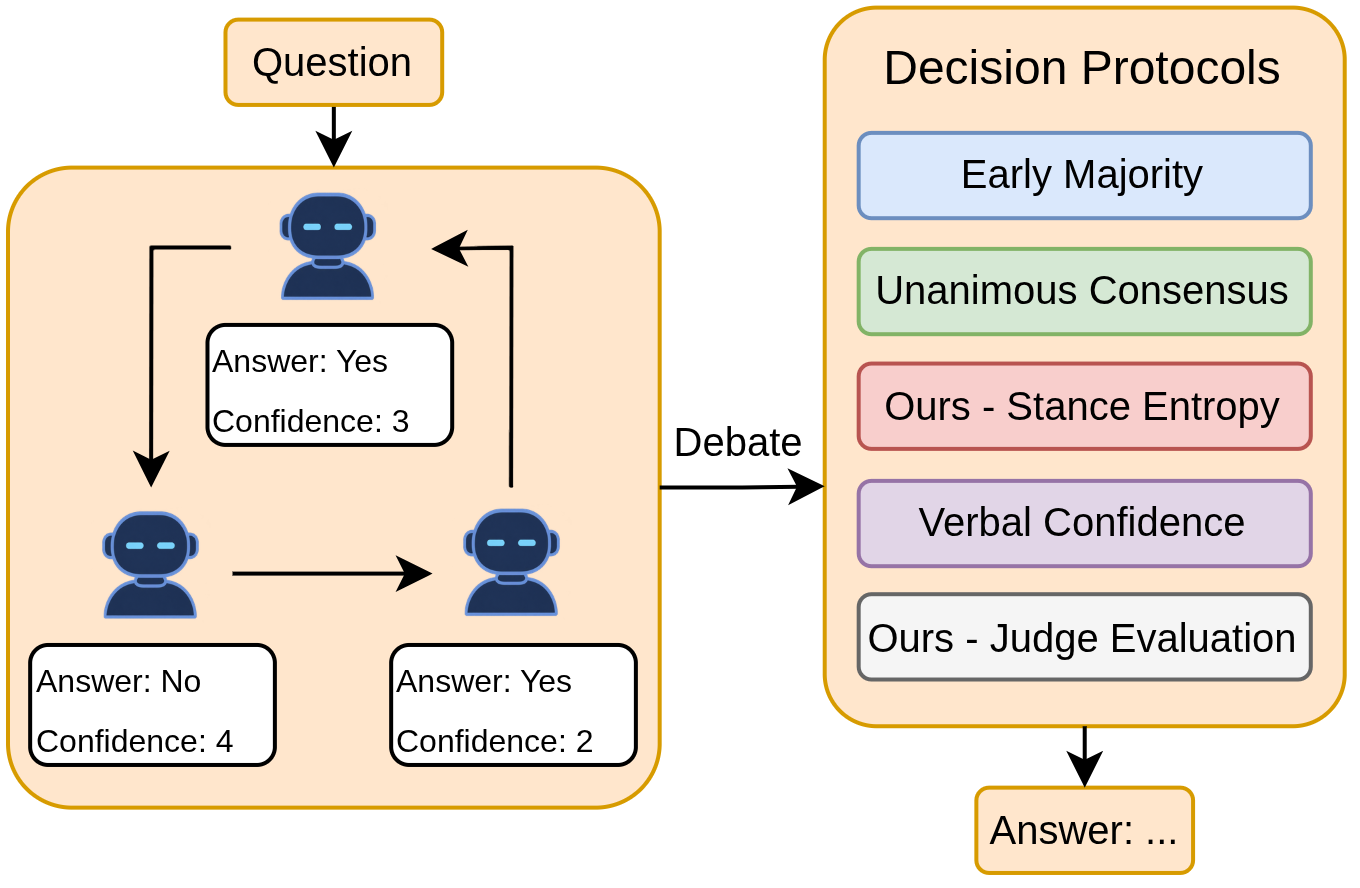}
                \caption{Our methodology for the experiments shown in a three-agent setting using a binary-choice dataset.}
                \label{fig:methodology}
    \end{figure}

     Employing judge agents backed by LLMs is an alternative to existing protocols \cite{smit2023should}. However, this further increases the computational overhead and is justified only if it increases the debate success.
     These observations motivate examining decision protocols to balance debate accuracy and computational efficiency. We aim to answer the following research questions (RQs):

    \begin{itemize}
    \item RQ1: How does MAD perform with different decision protocols compared to a single-agent approach across different tasks?
    \item RQ2: Which decision protocol provides greater scalability as the number of agents and candidate answers increase?
    \item RQ3: How does utilizing a judge agent to evaluate arguments compare to other decision protocols?
    \end{itemize}

    \paragraph{Contributions.} 
    Our contributions include: evaluating different agent configurations across multiple datasets and decision protocols, while identifying optimal protocols based on the task and debate setting; introducing stance entropy as an adaptation of Shannon entropy and determining optimal entropy thresholds for debate termination; introducing a judge-based evaluation method that adapts token usage to agent capabilities to derive the final decision while limiting additional token generation.
    
    

\section{\uppercase{Related Work}}

\paragraph{Decision Protocols in Multi-Agent Systems.}


MAD systems rely on specific debate termination and decision protocols. It has been shown that the choice of voting and consensus protocols influences the task performance \cite{kaesberg2025votingconsensus}. Prior work introduces DOWN, which bypasses MAD for high-confidence queries and triggers debate only under uncertainty \cite{eo2025debate}, ConfMAD, which incorporates explicit confidence expression, preventing premature convergence \cite{lin-hooi-2025-enhancing}, and RECONCILE, which introduces confidence-weighted voting \cite{chen2024reconcile}. 

Prior work emphasizes LLM-backed judges as decision-makers, introducing an adaptive stability-based stopping criterion to improve robustness and accuracy \cite{hu2025multi}.
Another work utilizes an LLM as a judge with a consensus-checking mechanism to improve efficiency \cite{Qian2025}. 
Other studies explicitly analyze the role of judges in resolving non-convergent debates \cite{liang-etal-2024-encouraging,xiong-etal-2023-examining}.
Collectively, these works demonstrate that LLM judges can go beyond aggregating agent votes and perform evaluative reasoning of debate context.
The increasing use of diverse personas and beliefs in agent systems poses challenges to the scalability of decision protocols. 
To tackle this challenge, we utilize an entropy-based and a judge-based protocol that adapts token usage to agent capability, and compare their performance with state-of-the-art protocols.



\paragraph{Entropy in Debate Dynamics.}


\textit{Entropy} quantifies how dispersed or concentrated a probability distribution is and has emerged as a measure to quantify uncertainty and disagreement in decision-making. In the context of debate dynamics, it provides a principled way to assess the degree of disagreement among agents' opinions. When agents converge on a single position, entropy remains low, reflecting high certainty within the group. Conversely, if agents adopt diverse positions, entropy increases, indicating greater uncertainty and disagreement. 
In particular, vote selection has been studied using entropy-based measures to quantify these aspects in decision-making \cite{gill2005entropy,argamon1999committee}. 
Prior work utilizes entropy in debate settings to assess agent confidence \cite{yoffe2024debunc}, to optimize LLM outputs \cite{chang2024evince,khalid2025ergo}, and to guide consensus formation \cite{duan2025enhancing}. While the application of entropy is increasing in multi-agent settings, less work is focused on leveraging it to determine when debates should terminate or to decide debate outcomes. Our work aims to fill this gap by modeling debate termination and outcome selection using stance entropy, an entropy-based method grounded in Shannon entropy. 
Unlike DOWN and ConfMAD, the introduced entropy protocol captures uncertainty at the multi-agent level by utilizing stance distribution rather than relying on the confidence of individual agents.

\section{\uppercase{Methodology}}

\subsection{Dataset}

We evaluate the generalizability of decision protocols across diverse tasks and voting patterns using four datasets: CUAD \cite{hendrycks2021cuad}, MMLU \cite{hendrycks2020measuring}, MMLU-Pro \cite{wang2024mmlu}, and StrategyQA \cite{geva2021did}. We distinguish between binary-choice (two answer choices) and multiple-choice (more than two answer choices) datasets. 
CUAD is a binary-choice dataset with Yes/No responses covering contract-related legal reasoning, while StrategyQA uses binary choices to evaluate multi-step commonsense and factual reasoning. MMLU and MMLU-Pro are multiple-choice question-answering datasets with four and ten options, respectively.
From MMLU, MMLU-Pro, and StrategyQA, we randomly selected 100 samples as debate questions. For the CUAD dataset, we randomly selected 100 samples from the clause types exclusivity, anti-assignment, revenue/profit sharing, minimum commitment, and license grant, balanced across labels. We retrieved the top 5 relevant passages from contracts using \texttt{text-embedding-3-large} \cite{openai2024embeddings} and FAISS \cite{johnson2019billion}.

\subsection{Debate Setting}

    We used a MAD framework in which each agent responds sequentially (Figure \ref{fig:methodology}). We tested three, five, and seven agents to examine the impact of increasing agent count while preventing structural disadvantages in majority voting due to tie outcomes in even-sized groups. Each turn, the agents utilized Chain-of-Thought (CoT) reasoning \cite{wei2022chain,kojima2022large}, output their arguments, a proposed answer, and a verbal confidence score for that answer. For the verbal confidence, we defined a Likert scale between 1 and 5, from very low to very high confidence. 
    
    We tested the homogeneous agents setting, where each agent is backed by the same LLM, and the heterogeneous agents setting, where each agent is backed by one of the three different LLMs. For the homogeneous agents setting, we utilized \texttt{Llama-3.1-8B-Instruct} \cite{dubey2024llama}, \texttt{Qwen2.5-7B-Instruct} \cite{qwen2}, \texttt{Phi-4-mini-instruct} \cite{abouelenin2025phi}, and \texttt{gpt-5-mini-2025-08-07} \cite{openai_gpt5_mini}, referred to as Llama-3.1, Qwen-2.5, Phi-4, and GPT-5-mini hereafter, respectively. In the heterogeneous agents setting, we excluded GPT-5-mini due to its relative strength and assigned Llama-3.1, Qwen-2.5, and Phi-4 to agents in a round-robin fashion.
    We used a temperature value of 0.7 for models that support temperature control to balance determinism and diversity in agent responses. 
    Each debate runs for up to five rounds, allowing agents to reach unanimous consensus on challenging samples, as observed in studies \cite{yin2023exchange,xiong-etal-2023-examining}.

    We compared the debate outcomes against the single agent, using the same CoT-reasoning approach as the agents in MAD. 
    The agents in MAD were limited to a maximum of five sentences per turn to prevent excessive verbosity across multiple rounds. 
    Agents were not instructed to optimize for token efficiency or minimize cost. 
    The decision protocols were evaluated on the same debates to ensure a more reliable comparison across protocols, leading to identical token usage for algorithms that do not terminate the debate. 
    Each experimental setting was run five times, with results averaged across runs.
    

\subsection{Decision Protocols}



\paragraph{Verbal Confidence.} 

The answer with the maximum average verbal confidence over all agents and rounds is selected as the final output. No early termination is applied. If two or more candidate answers share the same confidence scores, the final answer is selected randomly from among them.

\paragraph{Early Majority.}

The debate terminates at the end of any round in which an option attains a majority among the agents.
If no answer reaches the majority, the verbal confidence protocol is triggered.

\paragraph{Unanimous Consensus.}   

The debate terminates when all agents reach unanimous agreement on a single answer at the end of any round, which is then selected as the final output. If the agents do not reach unanimous consensus by the end of the debate, the early majority protocol is applied.


\paragraph{Stance Entropy.}   
The debate is terminated if the stance entropy is below the selected threshold value, in which case the most-supported answer is selected. If the entropy does not fall below the threshold until the end of the debate, the early majority protocol is applied.
We utilized the Shannon entropy \cite{shannon1948mathematical} to calculate the stance entropy.
In Equation~\ref{eq:stance_entropy}, $H$ is the stance entropy and is computed over valid agent outputs, $p_i$ is the probability of the $i$-th stance category in the full set of possible stances $V$, $H_\text{max}$ is the maximum possible entropy representing maximum uncertainty when all possible stances have equal probability, $n$ is the number of agents, $k$ is a scaling factor, and $\theta$ is the threshold entropy.

\begin{equation}\label{eq:stance_entropy}
\begin{cases}
\begin{aligned}
H &= -\sum_{i=1}^{|V|} p_i \log_2(p_i) \\
\theta &= H_\text{max} \cdot \left(1 - e^{-k n}\right)
\end{aligned}
\end{cases}
\end{equation}

\begin{figure}[htb!] 
          \centering
            \includegraphics[width=1.0\linewidth]{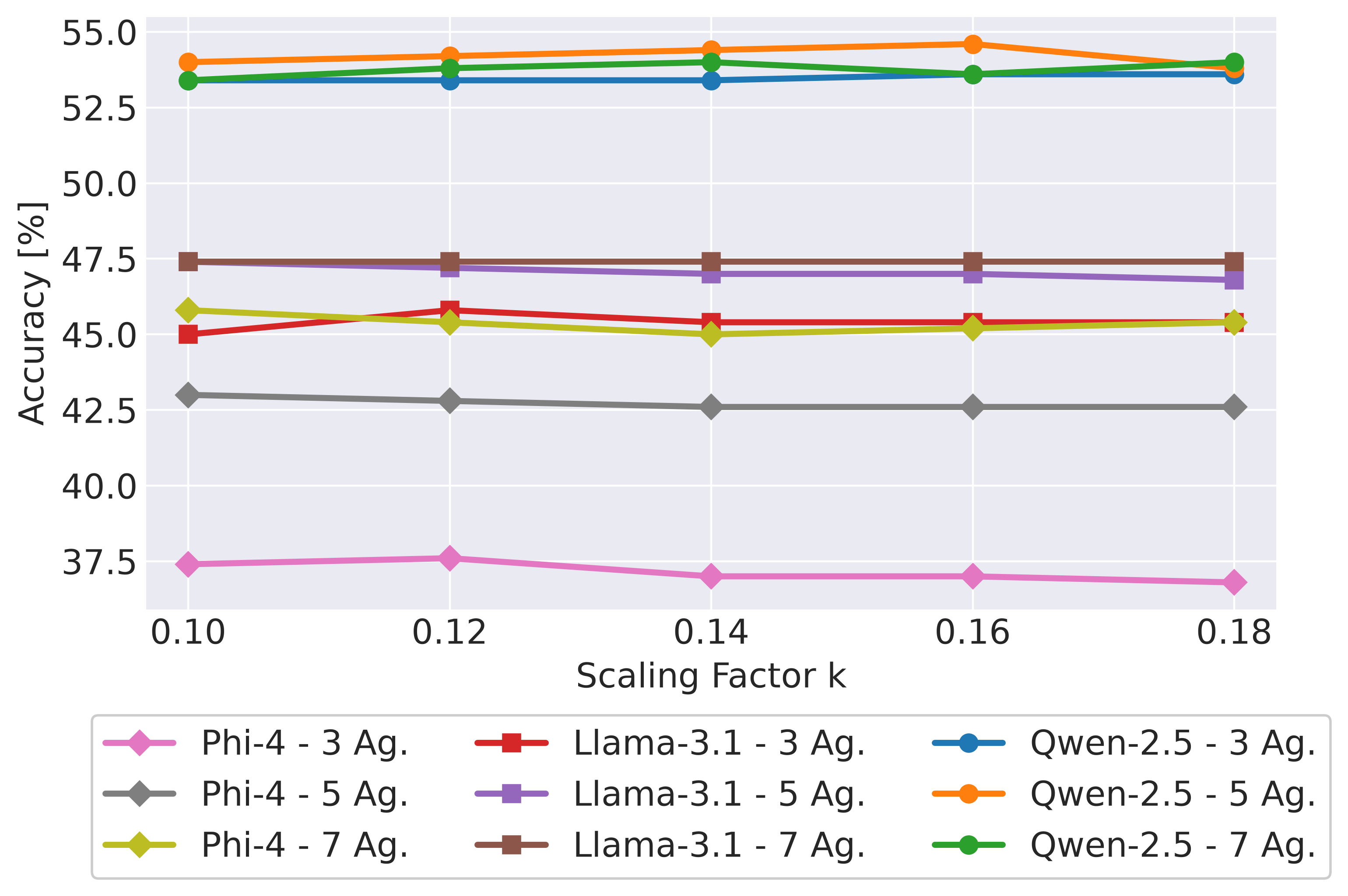}
            \caption{Mean accuracy (in \%) as a function of the scaling factor $k$ on MMLU-Pro across different debate settings. Ag. refers to the agent count in MAD.}
            \label{fig:sensitivity}
\end{figure}

\begin{figure}[htb!] 
          \centering
            \includegraphics[width=1.0\linewidth]{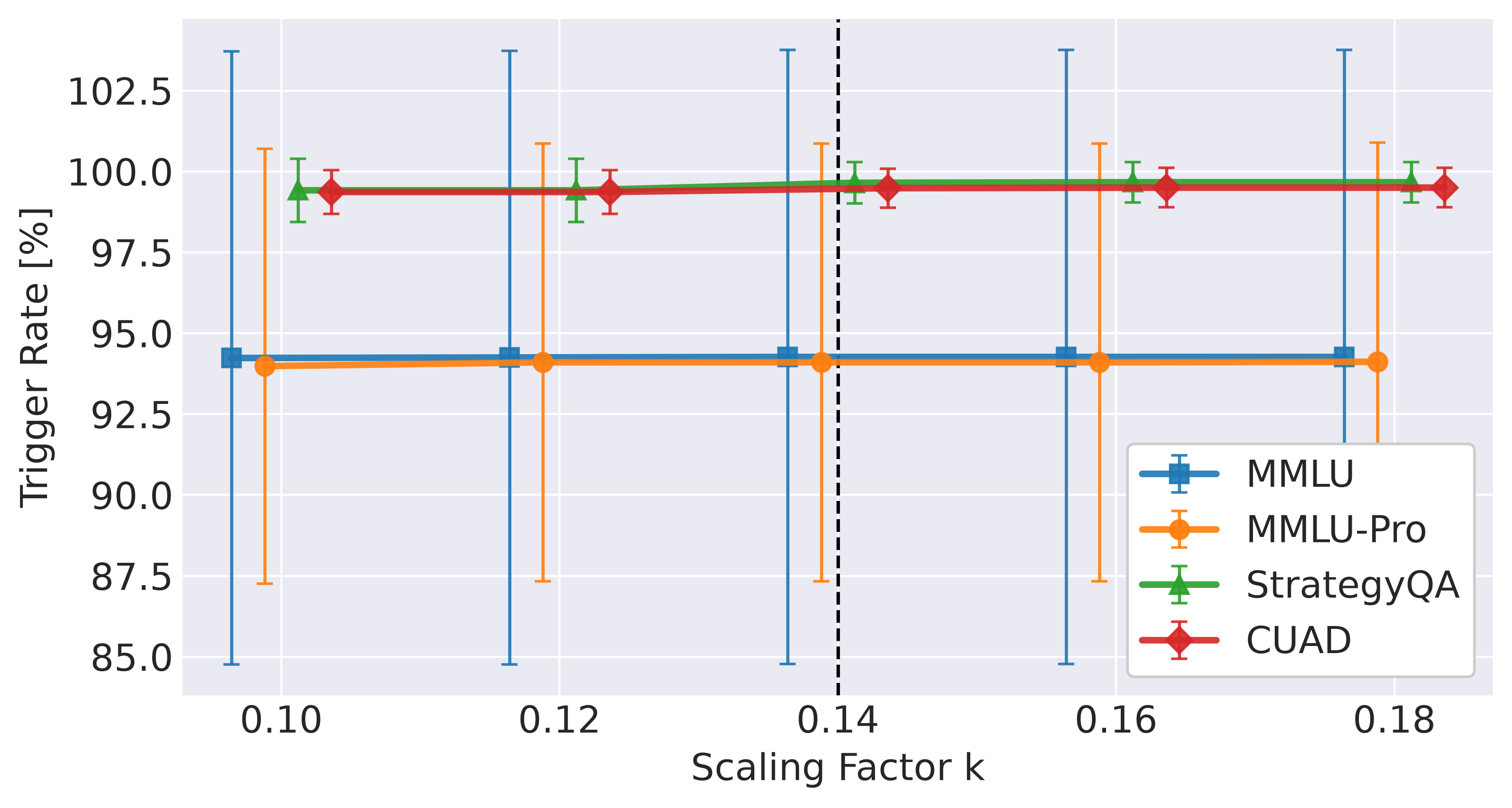}
            \caption{Mean trigger rate (in \%) as a function of the scaling factor $k$ on four datasets across different debate settings. Error bars show $\pm$1 standard deviation.}
            \label{fig:sensitivity2}
\end{figure}

To determine the scaling factor and the threshold value based on the number of agents, we used a separate set of 100 samples from the MMLU dataset and selected the value that yielded the highest average accuracy across different LLMs and numbers of agents. We hypothesize that this threshold generalizes well to other datasets with varying numbers of choices. Based on these experiments, we set the scaling factor to 0.14. 
This scaling factor represents an empirically calibrated operating point for four-option tasks.
It leads to different behaviors depending on the number of agents. With three agents on MMLU, a vote distribution of, e.g., two ``A'' votes and one ``C'' vote is insufficient to trigger termination and requires a unanimous consensus among agents. On the other hand, with seven agents, the scaling factor prevents termination in cases of diverse outputs such as three ``A'' votes, two ``B''  votes, and two ``C'' votes.
We performed a sensitivity analysis to evaluate the stability of accuracy under small variations of the scaling factor $k$. Figure \ref{fig:sensitivity} shows accuracy as a function of the scaling factor over the range [0.10, 0.18] on MMLU-Pro. The results indicate that accuracy varies smoothly and remains consistent across neighboring values, demonstrating stability and low sensitivity around the selected threshold.
Figure \ref{fig:sensitivity2} shows the mean trigger rate of stance entropy, with error bars representing the standard deviation, aggregated over Phi-4, Llama-3.1, Qwen-2.5, and GPT-5-mini and three-, five-, and seven-agent configurations. A clear split between binary-choice datasets and multiple-choice datasets is shown, where stance entropy triggers termination for multiple-choice datasets around $5.3\%$ less often than for the binary-choice datasets. 
This indicates that the trigger rate remains consistent across neighboring values and can generalize to datasets with a different number of answer candidates, while differentiating debate dynamics if the answer space affords disagreement among the agents.

\paragraph{Judge-Based Protocol.}



In this setting, we utilize \texttt{gpt-5-mini-2025-08-07} as the judge to evaluate agent arguments across eight dimensions: factual grounding, logical structure, evidence quality, clarity of reasoning, premise relevance, premise sufficiency, counterargument handling, and specificity. Each criterion is scored on a scale from 0 to 5, and the total score for each agent is computed as the sum of these eight dimensions, ranging from 0 to 40 points, inspired by prior work on multi‑dimensional argument quality assessment \cite{wachsmuth2017argumentation,wachsmuth2020intrinsic}.

At the end of Round 1, we first check for unanimous consensus. If all agents agree, the debate terminates with that answer without judge evaluation. This early check reduces computational cost when agents naturally converge. Otherwise, the judge evaluates each agent's argument and assigns quality scores. For each candidate answer, we compute the arithmetic mean score across agents proposing it and select the highest-scoring candidate. Agents then refine their arguments in Round 2, after which the judge repeats the evaluation. If the top candidates from Rounds 1 and 2 match, the debate terminates with that answer.
We hypothesize that this demonstrates the stability of the judge and the correctness of agents' reasoning. If the top candidates differ between the two rounds, or their scores are tied, we fall back on the early majority protocol for the first two rounds to resolve the conflict. 
With this design, the judge evaluates reasoning quality through argument analysis, while consensus mechanisms capture collective agreement among agents.

\subsection{Evaluation Method}

We evaluate the performance of decision protocols using the accuracy of the debate outcome and the generated token count. While accuracy evaluates task performance, the generated token count serves as a measure of debate efficiency, as decision protocols determine the length of the generated discourse. The generated token count refers to the average number of tokens generated per sample. 
We assessed statistical significance using paired t-tests over the accuracies of the five runs \cite{lakens2013calculating} and calculated effect sizes using Cohen's $d$ \cite{cohen2013statistical} based on the mean and standard deviation of the paired differences.
For the homogeneous-agents setting, the baseline was the single-agent condition using the same underlying LLM. For the heterogeneous-agents setting, the baseline was the single-LLM condition using the model that achieved the highest accuracy among the LLMs in the heterogeneous setting, namely Qwen-2.5.

\section{\uppercase{Results and Discussion}}

\subsection{RQ1: Accuracy across Protocols}


\begin{figure}[htb!] 
          \centering
            \includegraphics[width=1.0\linewidth]{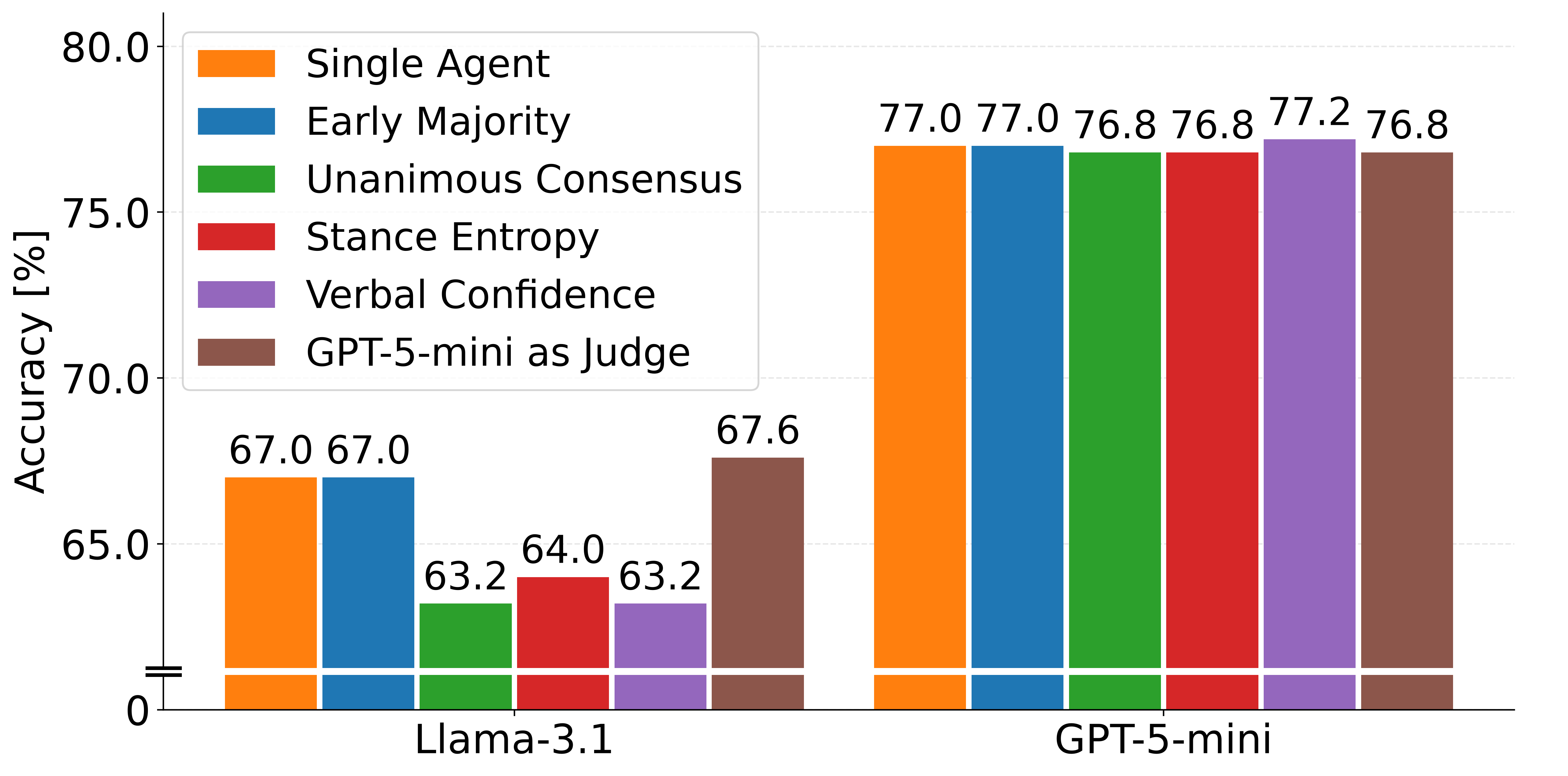}
            \caption{Accuracy (in \%) across decision protocols for the StrategyQA dataset and three-agent setting.}
            \label{fig:plot2}
\end{figure}

We hypothesize that the effectiveness of MAD depends on the decision protocol and task characteristics, and that protocols capable of effectively aggregating diverse agent perspectives can improve task accuracy.
For CUAD, while MAD achieves similar performance across protocols, single-agent performance is higher (e.g., early majority achieves the highest accuracy in the three-agent setting, at 71.0\% and 80.6\% for Qwen-2.5 and GPT-5-mini, respectively, compared to 72.8\% and 83.4\% for single-agent).
Figure \ref{fig:plot2} shows the accuracy for StrategyQA in the three-agent setting.
MAD, particularly with early majority, delivers competitive performance compared to the single-agent approach. 
Phi-4 was an exception, whose substantially lower single-agent accuracy may reflect its weaker multi-hop reasoning. Among the decision protocols, early majority delivers the best overall performance independent of LLMs. 
Stance entropy excludes invalid answers from its calculation, whereas unanimous consensus requires all answers to be valid, explaining the accuracy difference.

\begin{figure}[htb!] 
          \centering
            \includegraphics[width=1.0\linewidth]{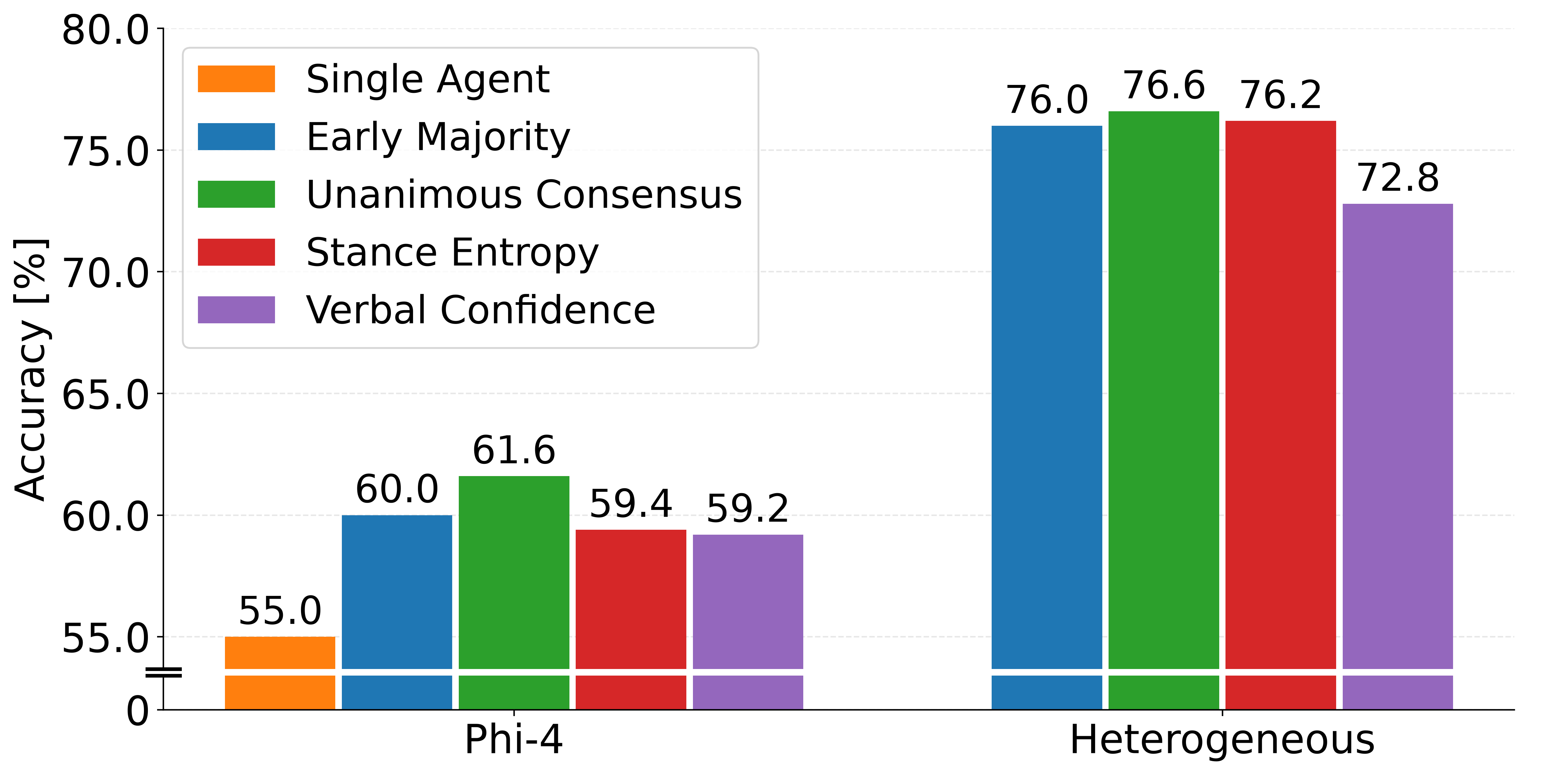}
            \caption{Accuracy (in \%) across decision protocols for the MMLU and five-agent setting. The single-agent configuration does not apply to the heterogeneous setting.}
            \label{fig:plot3}
\end{figure}

\begin{figure}[htb!] 
          \centering
            \includegraphics[width=1.0\linewidth]{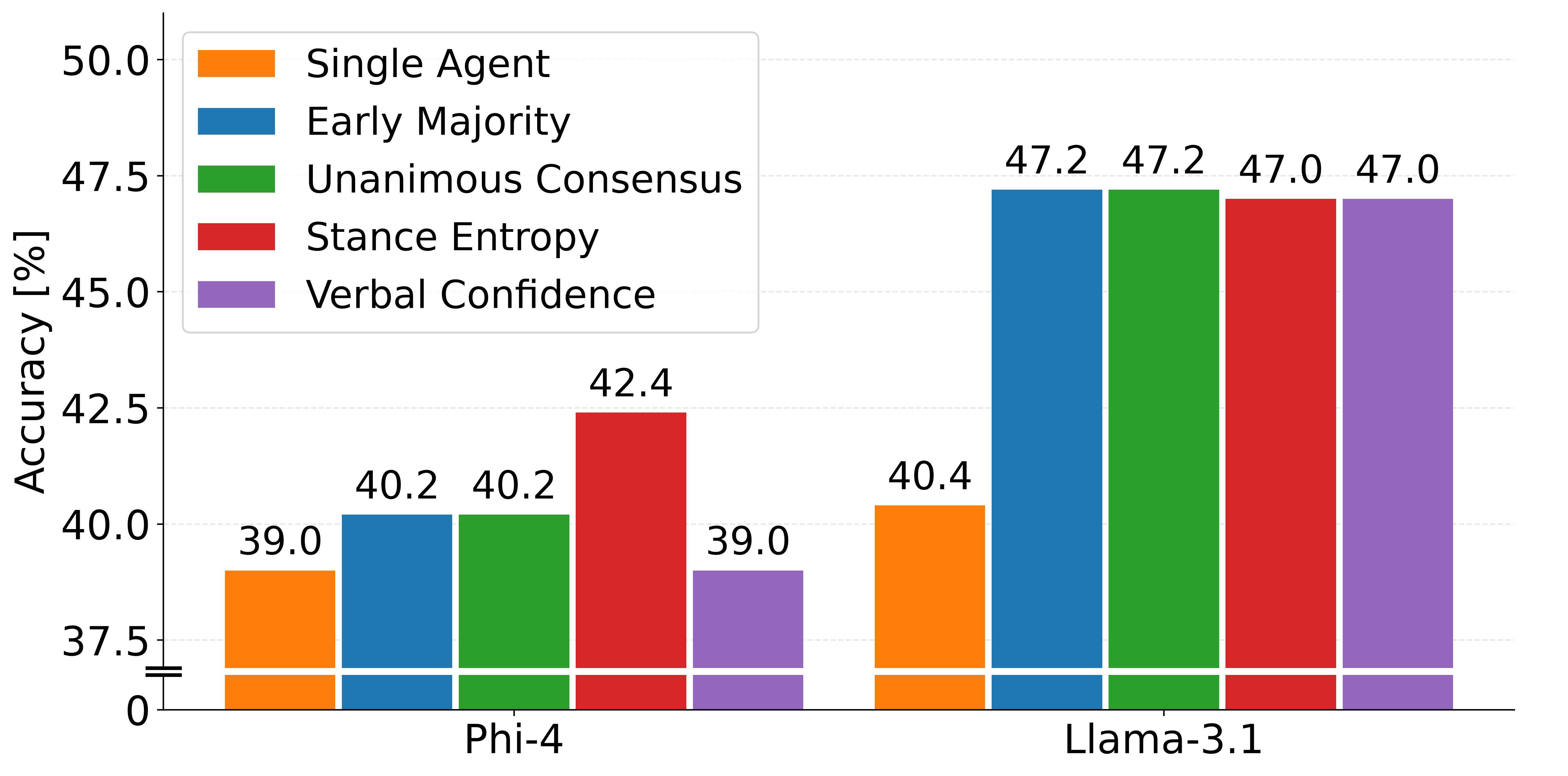}
            \caption{Accuracy (in \%) across decision protocols for the MMLU-Pro dataset and five-agent setting.}
            \label{fig:plot4}
\end{figure}

Figures \ref{fig:plot3} and \ref{fig:plot4} show the accuracy for the MMLU and MMLU-Pro with five agents, respectively. In both datasets, the single-agent performance is weaker, while for MMLU the unanimous consensus shows the strongest performance (61.6\%, $d$=2.53, $p<$0.01 and 76.6\%, $d$=2.67, $p<$0.01 for Phi-4 and the heterogeneous setting, respectively) for the five-agent setting. When the number of candidate answers increases, as in the MMLU-Pro, the strongest performance for Phi-4 is observed for stance entropy (42.4\%, $d$=0.50, $p>$0.10). For Llama-3.1, the decision protocols show similar performance, while early majority (47.2\%, $d$=2.45, $p<$0.01) and unanimous consensus (47.2\%, $d$=1.57, $p<$0.05) result in the highest accuracy. 

The first research question examined how decision protocols influence debate outcomes compared to the single-agent approach.
For StrategyQA and CUAD, early majority provides balanced performance but faces competition from the single-agent approach, which achieves high accuracy on multi-step and legal reasoning tasks.
For multiple-choice datasets, depending on the number of answer candidates and feasibility of full agreement, stance entropy and unanimous consensus are the best candidates.

\subsection{RQ2: Scalability of the Protocols} \label{sec:rq2}

\begin{figure}[htb!] 
          \centering
            \includegraphics[width=1.0\linewidth]{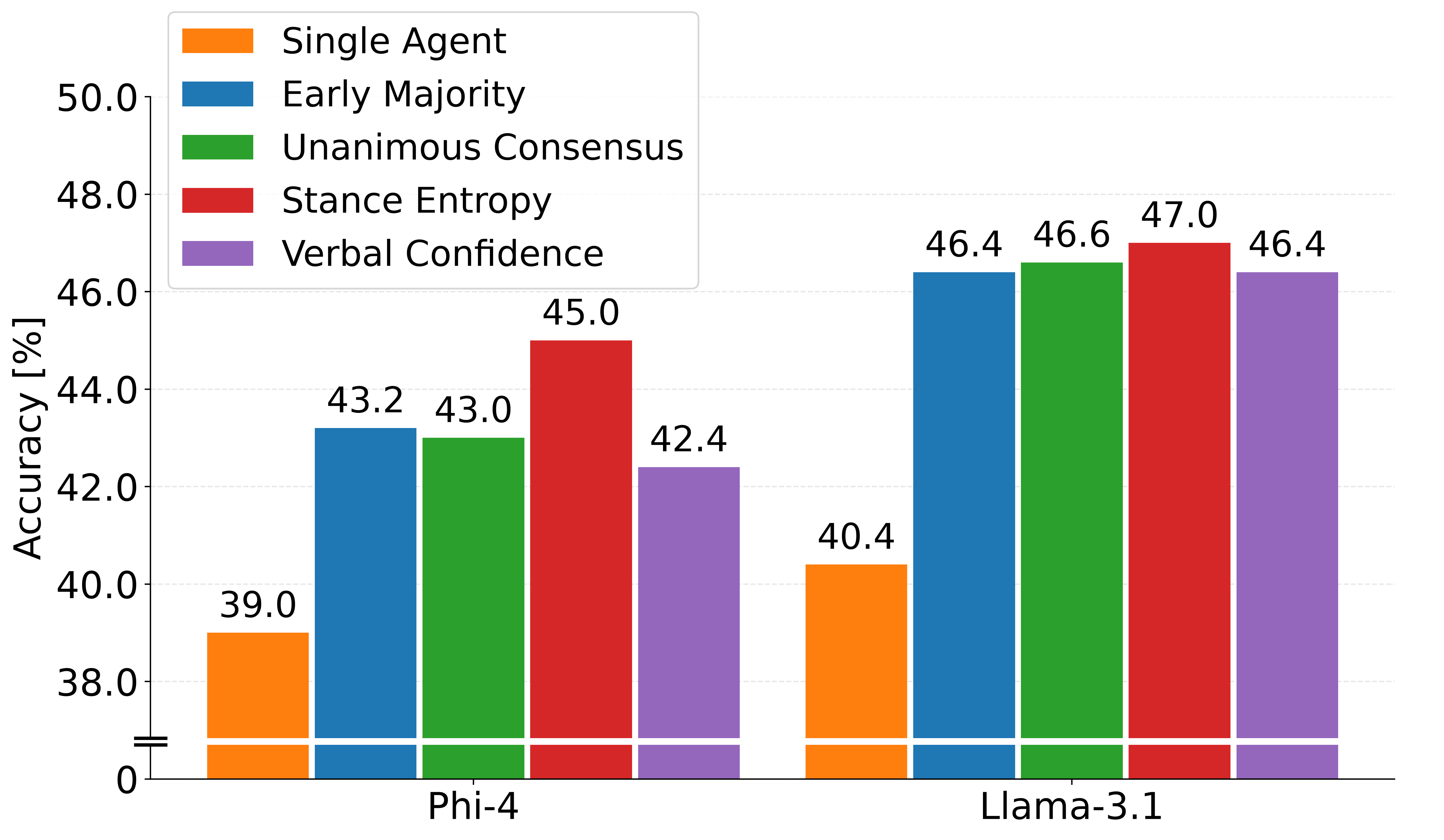}
            \caption{Accuracy (in \%) across decision protocols for the MMLU-Pro dataset and seven-agent setting.}
            \label{fig:plot5}
\end{figure}

\begin{figure}[htb!] 
          \centering
            \includegraphics[width=1.0\linewidth]{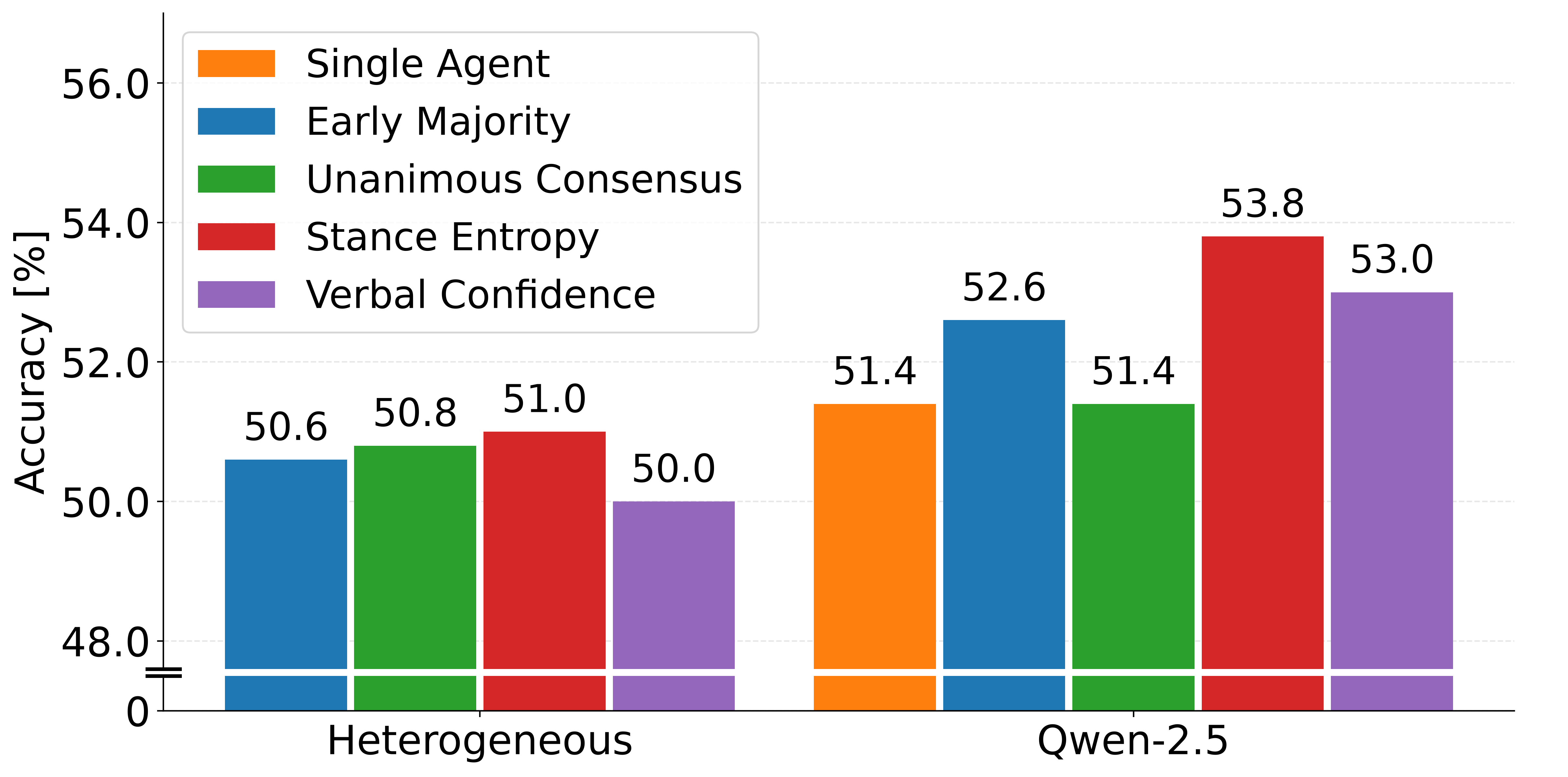}
            \caption{Accuracy (in \%) across decision protocols for the MMLU-Pro and seven-agent setting. The single-agent configuration does not apply to the heterogeneous setting.}
            \label{fig:plot6}
\end{figure}

\begin{figure}[htb!] 
          \centering
            \includegraphics[width=1.0\linewidth]{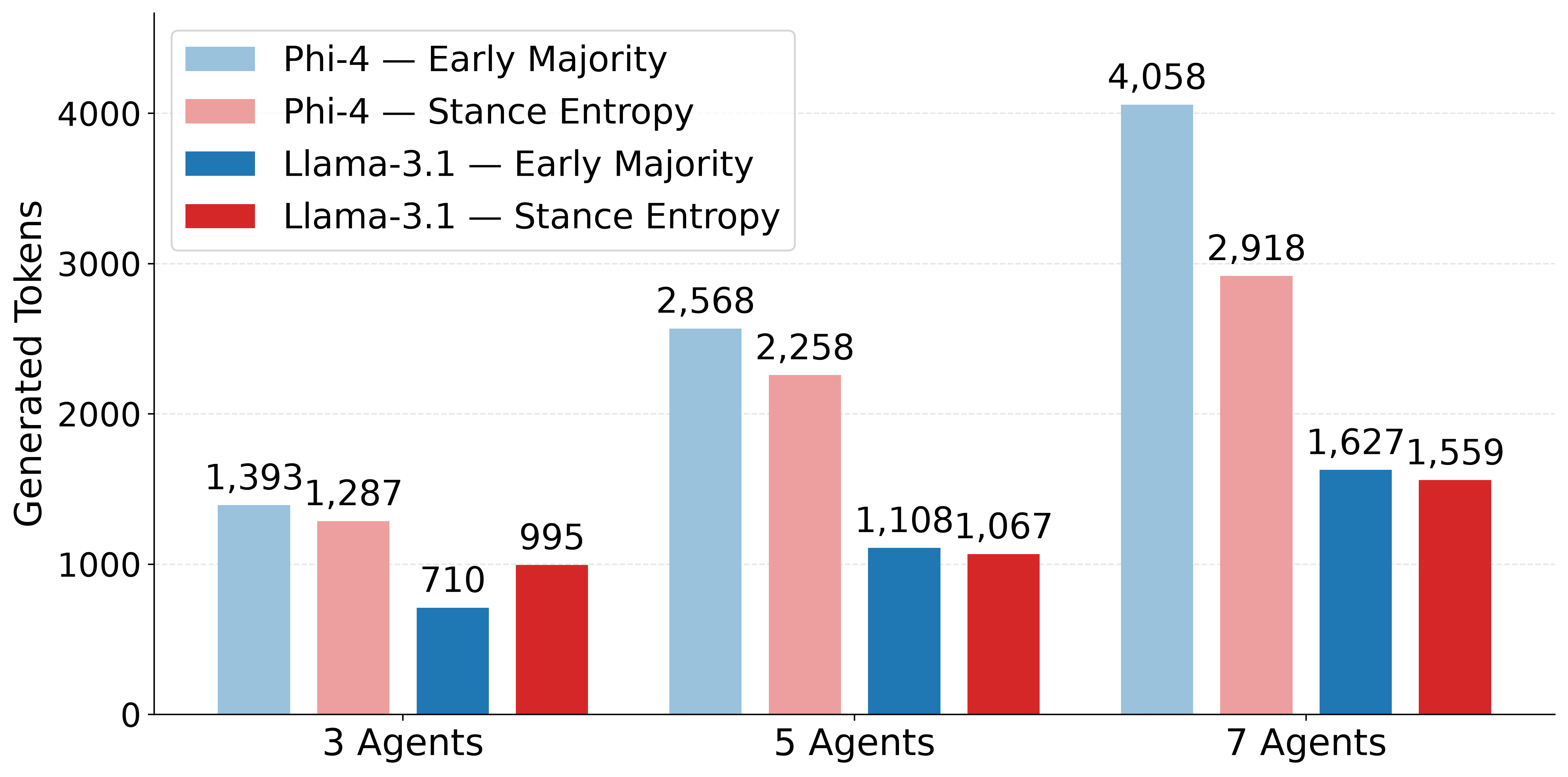}
            \caption{Generated token count for early majority and stance entropy in comparison for the MMLU dataset and different agent numbers in the debate.}
            \label{fig:plot7}
\end{figure}

\begin{figure}[htb!] 
          \centering
            \includegraphics[width=1.0\linewidth]{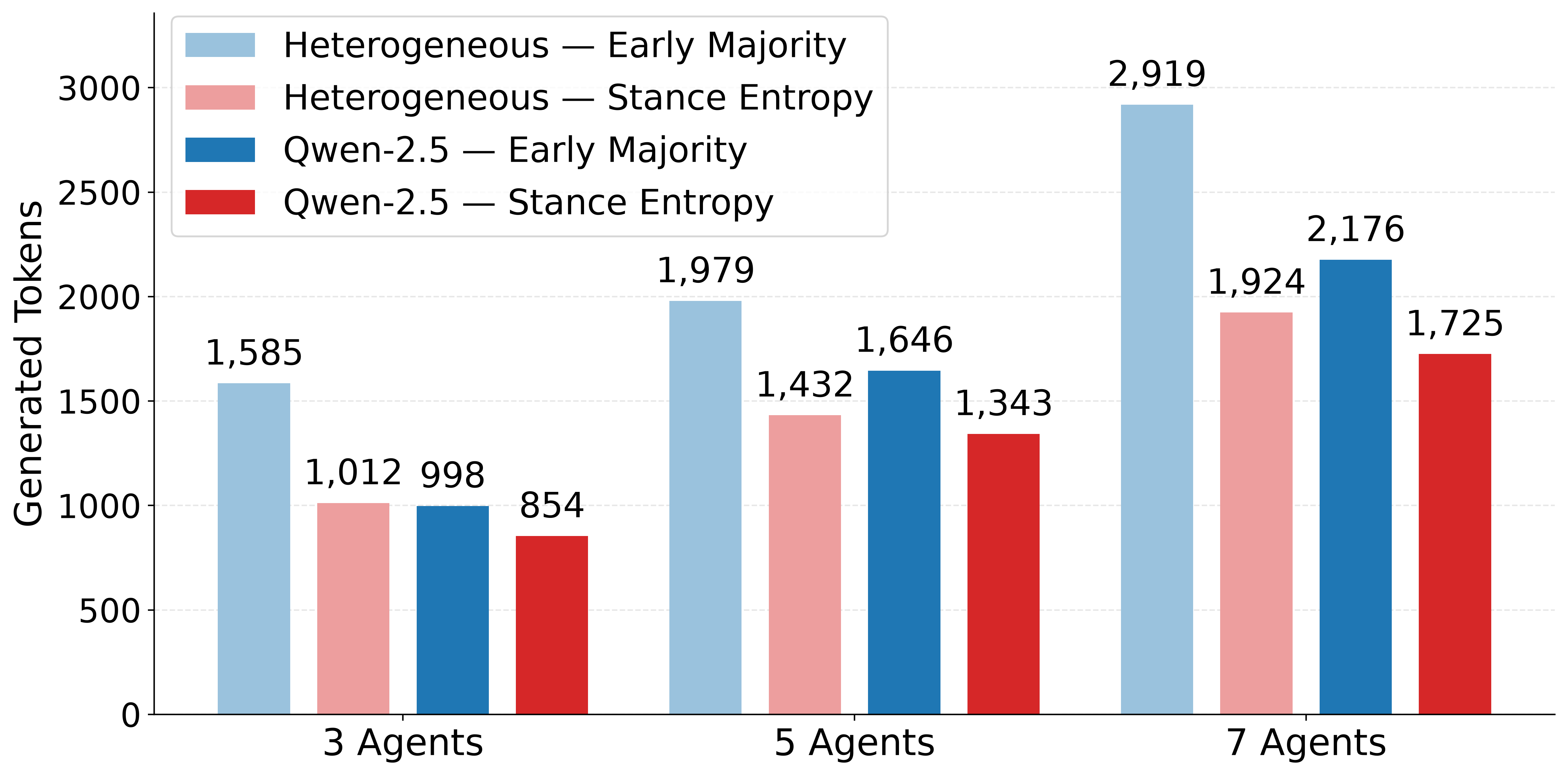}
            \caption{Generated token count for early majority and stance entropy in comparison for the MMLU-Pro dataset and different agent numbers in the debate.}
            \label{fig:plot8}
\end{figure}

We investigated the scalability of decision protocols as the number of agents and candidate answers increases. 
Figures \ref{fig:plot5} and \ref{fig:plot6} show the accuracy for the seven-agent setting. Stance entropy yields higher accuracy than other protocols (45.0\%, $d$=0.97, $p<$0.10; 47.0\%, $d$=1.96, $p<$0.05; 51.0\%, $d$=-0.08, $p>$0.10; and 53.8\%, $d$=1.23, $p<$0.10 for Phi-4, Llama-3.1, heterogeneous setting, and Qwen-2.5, respectively), suggesting that is a more scalable protocol to higher number of agents, while the single agent can not reach to the accuracy of MAD except the Qwen-2.5, where it shows the same accuracy as unanimous consensus (51.4\%), but remain weak compared to the other protocols.
The verbal confidence lacks early termination, while unanimous consensus requires more rounds, resulting in higher token usage. Thus, we compare token counts only for early majority and stance entropy.
Figures \ref{fig:plot7} and \ref{fig:plot8} compare these protocols for MMLU and MMLU-Pro datasets, respectively. In general, the stance entropy exhibits the lowest generated token usage for multiple-choice datasets.

The second research question investigated which decision protocol is most scalable as the number of agents and candidate answers increase.
For stance entropy, we framed the debate decision as an entropy minimization problem over the stance distribution of agents and provided a grounded alternative to existing protocols. Instead of using entropy solely to quantify disagreement, our approach uses it as a control mechanism for terminating debates and making final decisions in multi-agent debates.
We observed that stance entropy achieves higher accuracy with seven agents, while maintaining lower token usage, particularly on multiple-choice datasets, showing better scalability.

\subsection{RQ3: Effect of Judge}

\begin{figure}[htb!] 
          \centering
            \includegraphics[width=1.0\linewidth]{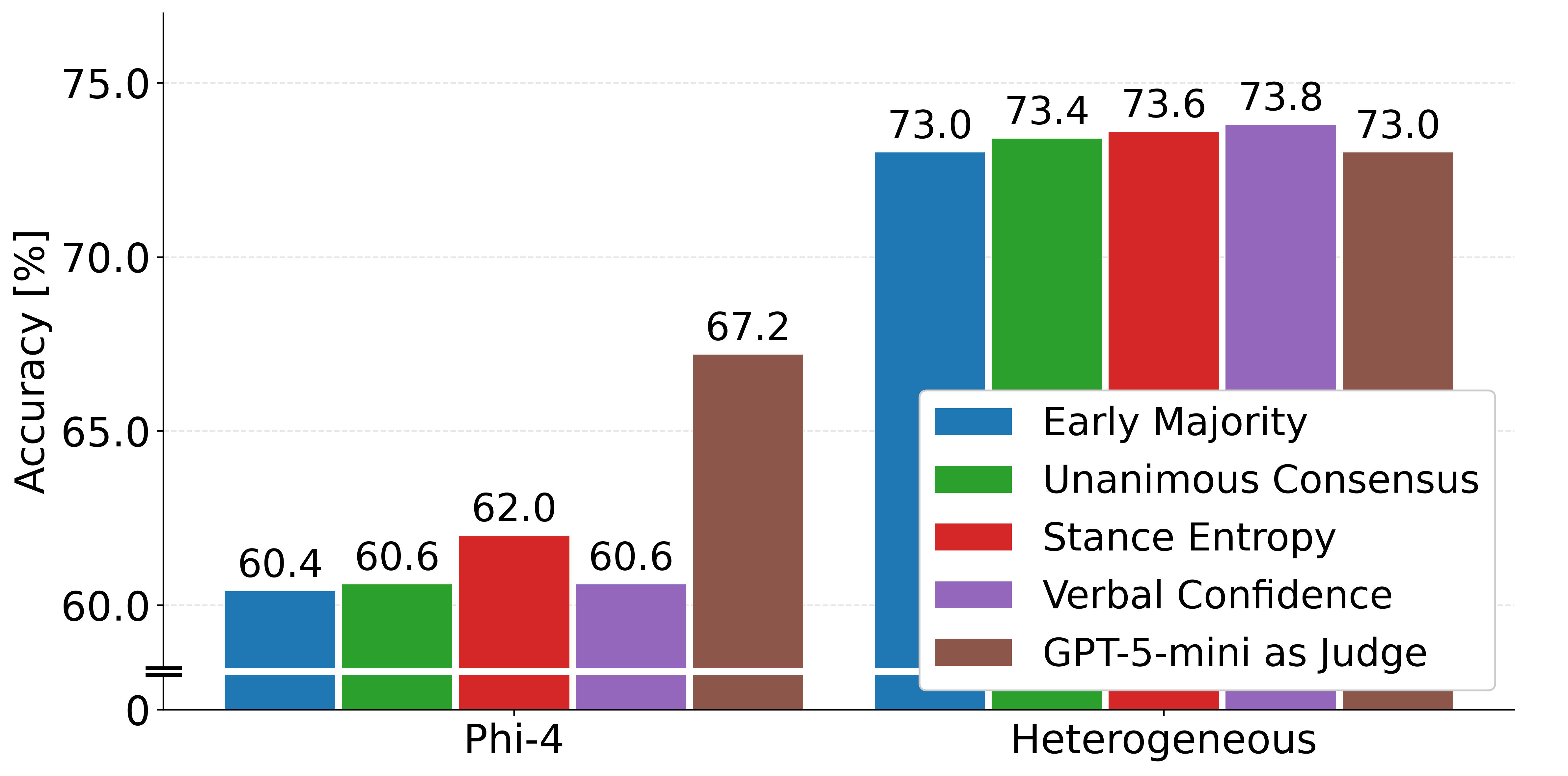}
            \caption{Accuracy (in \%) across decision protocols for the MMLU dataset and three-agent setting.}
            \label{fig:plot9}
\end{figure}

\begin{figure}[htb!] 
          \centering
            \includegraphics[width=1.0\linewidth]{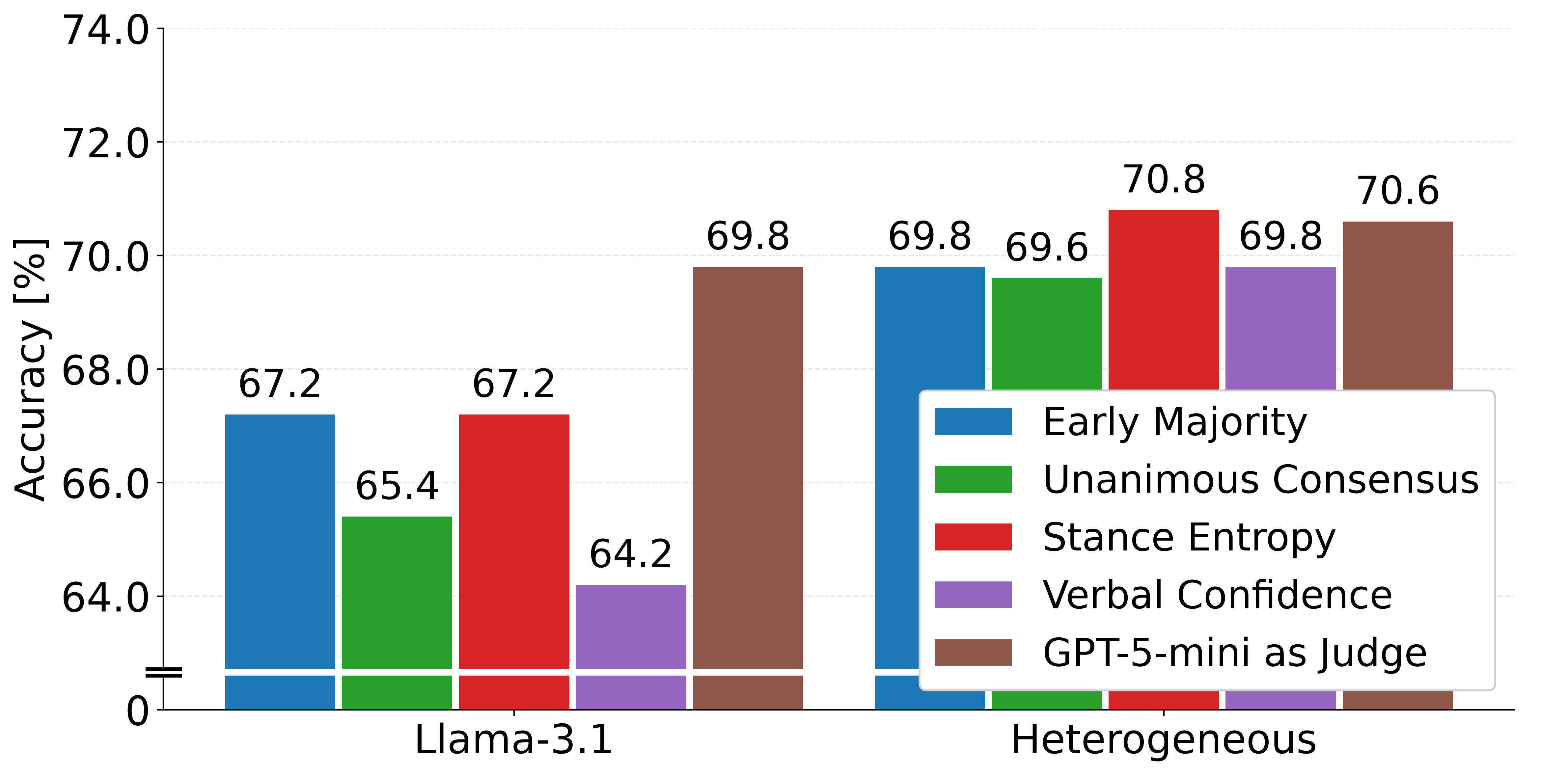}
            \caption{Accuracy (in \%) across decision protocols for the CUAD dataset and three-agent setting.}
            \label{fig:plot10}
\end{figure}

Figures \ref{fig:plot9} and \ref{fig:plot10} show the accuracy for the three-agent setting for MMLU and CUAD datasets, respectively.
For MMLU and MMLU-Pro with the judge evaluation, we achieved the highest accuracy in smaller models, such as Phi-4 (67.2\%, $d$=3.07, $p<$0.01 for MMLU), as expected due to the strong LLM as judge. The judge was able to identify the arguments that provided the strongest support for the correct answer based on argument quality.
For StrategyQA and CUAD with Phi-4 and Llama-3.1, the judge evaluation outperforms the other protocols in the three-agent setting. However, performance improvements appeared to depend on particular debate settings such as the strength of LLMs or agent counts, rather than reflecting a generalizable improvement. 
Overall, on MMLU and MMLU-Pro datasets, this setting achieved the highest accuracy across all LLM configurations. On MMLU-Pro, it could increase the accuracy of MAD backed by GPT-5-mini and outperform single-agent and other decision protocols.


The third research question analyzed the judge-based protocol in comparison with other protocols. 
Together with its fallback mechanism to majority voting, it combines argument-quality assessment with agent voting, providing a hybrid approach for decision-making. 
For multiple-choice datasets, it improves the accuracy in the five-agent setting; but the improvement diminishes for legal and multi-step reasoning tasks or in debates involving stronger models.

\section{\uppercase{Conclusion}}\label{sec:conclusion}

We systematically investigated decision protocols in MAD and highlighted their strengths and limitations under different debate settings and tasks. 
Our results show that the single-agent method can be competitive, especially if the decision mechanism is not appropriately matched to the scale of the debate or characteristics of the task. 
The scalability of MAD depends on how agent perspectives are aggregated as the number of agents increases. 
In particular, the proposed stance entropy leads to a competitive balance between accuracy and computational efficiency, especially in debates with a higher number of agents and if there are more than two answer candidates. The judge-based evaluation improves accuracy, especially in multiple-choice datasets, if the judging process is applied selectively.
Overall, adaptive decision protocols are key to robust and reliable MAD systems. For easy applicability, we provide access to our source files.
\footnote{https://github.com/Advancing-Machine-Human-Reasoning-Lab/protocols}


\bibliographystyle{apalike}
{\small
\bibliography{custom}}

\appendix

\section{\uppercase{Appendix}}

\subsection{Agent Instructions}
\label{sec:agent_prompt}

Figure \ref{fig:agent_prompt_approach} shows the prompting approach for debates in the MMLU and MMLU-Pro examples, with Agents 1 and 2 participating in the debate context and Agent 3 providing the next response. The answer format is ``Yes/No'' for datasets with binary labels (StrategyQA and CUAD) and ``A/B/C/D'' or ``A/B/C/D/E/F/G/H/I/J'' for multiple-choice datasets (MMLU and MMLU-Pro, respectively). Prompts for StrategyQA and CUAD do not include the subject or answer choices. In CUAD, the system prompt additionally provides the clause type and the relevant contract passages for each sample.
Figure~\ref{fig:judge_prompt} shows the judge evaluation prompt that implements a structured eight-dimensional rubric for assessing reasoning quality. The judge evaluates each agent's contribution, assigns a score for each criterion, and extracts each agent's final answer.

\begin{figure*}[hb!]
\centering
\begin{tikzpicture}
  \node[draw, rectangle, rounded corners=5pt, fill=lightgray!20] {\parbox{15.5cm}{
    \textbf{System:} You are \{Name\}. Confidence Scale:\\
    \{Confidence Scale\} \\
    \\
    Engage critically in an active debate to determine the correct answer. Contribute your reasoning in no more than five sentences. Finalize your answer with the choice you believe is most appropriate (\{Valid Options\}) and provide your confidence level. 
    Provide your decision in the following format:\\
    Answer: $<$\{Valid Options\}$>$; Confidence: $<$N$>$\\
    \\
    \textbf{User:} Agent 1: ...\\
    Agent 2: ...\\
    \\
    \#\#\#\\
    \\
    Subject: \{Subject\}\\
    Question: \{Question\}\\
    Choices: \\
    \{Answer Choices\}\\
    Let's think step by step.\\
    \\
    \textbf{Assistant:} {\{Answer\}}
  }};
\end{tikzpicture}
\caption{The prompting approach for the MMLU and MMLU-Pro datasets, demonstrating Agent 3 as the designated respondent in the first round.}
\label{fig:agent_prompt_approach}
\end{figure*}

\begin{figure*}[htb!]
\centering
\begin{tikzpicture}
  \node[draw, rectangle, rounded corners=5pt, fill=lightgray!20] {\parbox{15.5cm}{
    \textbf{System:} You are an expert debate judge evaluating reasoning quality.\\
    \\
    \textbf{User:} You are evaluating the quality of reasoning in a multi-agent debate. Score each agent on EIGHT dimensions (0-5 scale each):\\
    \\
    \textbf{1. Factual Grounding (0-5):} 5 = Multiple specific facts, all verifiable and accurate; 4 = Several facts, mostly verifiable; 3 = Some facts but mixed with generalizations; 2 = Mostly vague, few specific facts; 1 = Largely unsupported claims; 0 = False or contradictory statements\\
    \\
    \textbf{2. Logical Structure (0-5):} 5 = Rigorous reasoning chain, all steps justified; 4 = Clear reasoning with minor gaps; 3 = Partially logical, some leaps; 2 = Weak connections between claims; 1 = Mostly assertions without justification; 0 = Incoherent reasoning\\
    \\
    \textbf{3. Evidence Quality (0-5):} 5 = Specific citations, precise examples, quantitative data; 4 = Good specific examples; 3 = Some examples but generic; 2 = Vague references; 1 = Minimal evidence; 0 = No supporting evidence\\
    \\
    \textbf{4. Clarity of Expression (0-5):} 5 = Exceptionally clear, well-organized argument; 4 = Clear with good structure; 3 = Understandable but somewhat disorganized; 2 = Confusing structure or unclear points; 1 = Hard to follow; 0 = Incomprehensible\\
    \\
    \textbf{5. Premise Relevance (0-5):} 5 = All premises directly support the conclusion; 4 = Most premises relevant; 3 = Some relevant premises; 2 = Many irrelevant points; 1 = Mostly irrelevant; 0 = No relevant premises\\
    \\
    \textbf{6. Premise Sufficiency (0-5):} 5 = Comprehensive coverage of necessary points; 4 = Covers most necessary points; 3 = Covers some key points; 2 = Missing several key points; 1 = Severely incomplete; 0 = Insufficient premises\\
    \\
    \textbf{7. Counterargument Handling (0-5):} 5 = Addresses and refutes counterarguments thoroughly; 4 = Addresses main counterarguments; 3 = Acknowledges some counterarguments; 2 = Minimal engagement with opposing views; 1 = Ignores counterarguments; 0 = Unaware of counterarguments\\
    \\
    \textbf{8. Specificity (0-5):} 5 = Highly specific with precise details; 4 = Good level of detail; 3 = Moderately specific; 2 = Mostly general statements; 1 = Very vague; 0 = No specifics\\
    \\
    \textbf{Output Format:}\\
    Agent $i$: - Factual Grounding: [0-5], - Logical Structure: [0-5], - Evidence Quality: [0-5], - Clarity of Expression: [0-5], - Premise Relevance: [0-5], - Premise Sufficiency: [0-5], - Counterargument Handling: [0-5], - Specificity: [0-5], - Total Score: [sum of above, max 40], - Agent's Answer: [extract answer]\\
    \\
    \textbf{IMPORTANT:} Agent's Answer must be EXACTLY one of: \{Answer Format\}. Extract from agent's statement. If unclear, use ``Unknown''.\\
    \\
    Round \{r\} Discussion: \{Agent outputs\}\\
    \\
    \textbf{Assistant:} \{Judge Evaluation\}
  }};
\end{tikzpicture}
\caption{Eight-dimensional judge evaluation prompt used to assess agent reasoning quality. The judge outputs scores (0-5) for each dimension and extracts the agent's final answer.}
\label{fig:judge_prompt}
\end{figure*}

\subsection{Confidence Scale}

The confidence scale in Table \ref{tab:confidence_scale} is used by the agents to evaluate their own confidence in their answers.
We defined the scale and refined it through discussion, using ChatGPT to enhance the clarity of the descriptions \cite{chatgpt2025}.

\begin{table*}[htb!]
    \caption{The confidence scale and its description for the verbal confidence decision protocol.}
    \centering
    \begin{tabular}{ l | l }
        \hline 
        Score & Description \\
        \hline
        \multirow{4}{*}{1 (Very Low)} 
        & You have no reliable knowledge or relevant information to answer this question. \\
        & You might attempt to explore the problem, but quickly realize you lack the \ \\
        & foundational understanding needed. You are essentially guessing with no logical \\
        & basis. It's okay to admit when you don't know. \\
        \hline
        \multirow{3}{*}{2 (Low)} 
        & You have minimal understanding or minimal information. You might have some \\  
        & initial intuition or weak evidence, but it's insufficient to make any firm decision.  \\
        & Your answer is highly tentative, and you feel uncertain about its validity. \\
        \hline        
        \multirow{4}{*}{3 (Moderate)} 
        & You have a partial understanding with some supporting evidence or reasoning. \\
        & You can construct a reasonable argument, but significant doubts remain about \\
        & whether your answer is correct. You see the logic but recognize notable gaps \\
        & or alternative possibilities. \\
        \hline
        \multirow{4}{*}{4 (High)} 
        & You have strong logical backing and clear reasoning that support your answer. \\   
        & The evidence and your analysis align well, and you believe the answer is likely \\
        & correct. You acknowledge the possibility of minor errors but feel confident in \\
        & your conclusion. \\
        \hline  
        \multirow{4}{*}{5 (Very High)} 
        & You have thoroughly evaluated the problem with comprehensive reasoning and \\
        & unmistakable evidence. You solved it without significant difficulties and cannot \\
        & identify any flaws in your logical analysis. You are entirely sure the answer is \\
        & correct. \\
        \hline 
    \end{tabular}
    \label{tab:confidence_scale}
\end{table*}

\subsection{Experimental Results}
\label{sec:results_tables}

Tables \ref{tab:phi4_all_terminations}, \ref{tab:llama-3.1_all_terminations}, \ref{tab:qwen-2.5_all_terminations}, \ref{tab:gpt-5-mini_all_terminations}, and \ref{tab:heterogeneous_all_terminations} show the accuracy (in \%) with standard deviation and average generated token count per sample for each dataset, number of agents in debates, and decision protocols, with single-LLM results included for the corresponding models. Bold values highlight the best result except the judge-based evaluation, while bold values marked with a star ($^{\star}$) denote the highest accuracy, including the judge-based evaluation. 

\begin{table*}[htb!]
\caption{Phi-4: Debate accuracy (in \%) with standard deviation and average generated token count per sample across decision protocols.
Bold values indicate the best results except the judge-based evaluation, while values marked with $^{\star}$ denote the highest accuracy, including the judge-based evaluation. \#Ag. denotes the number of agents.}
\centering
\setlength{\tabcolsep}{3pt} 
\begin{tabular}{l | c | cc | cc | cc | cc }
\hline
\multirow{2}{*}{Protocol} & \multirow{2}{*}{\#Ag.}
& \multicolumn{2}{c|}{MMLU}
& \multicolumn{2}{c|}{MMLU-Pro}
& \multicolumn{2}{c|}{StrategyQA}
& \multicolumn{2}{c}{CUAD} \\
& 
& Accuracy & Token
& Accuracy & Token
& Accuracy & Token
& Accuracy & Token \\
\hline
Single Agent & 1 & 55.0 $\pm$ 2.3 & 269 & 39.0 $\pm$ 3.5 & 395 & 59.2 $\pm$ 3.8 & 172 & 64.8 $\pm$ 3.3 & 224 \\
\hline
\multirow{3}{*}{\makecell{Early\\Majority}}
& 3 & 60.4 $\pm$ 4.3 & 1393 & 36.4 $\pm$ 2.9 & 1914 & 67.6 $\pm$ 2.7 & 387 & 64.4 $\pm$ 2.3 & 1117 \\
& 5 & 60.0 $\pm$ 2.0 & 2568 & 40.2 $\pm$ 3.0 & 3192 & 66.8 $\pm$ 3.8 & 706 & \textbf{65.2$^{\star}$ $\pm$ 1.3} & 2064 \\
& 7 & 59.2 $\pm$ 5.4 & 4058 & 43.2 $\pm$ 1.8 & 4801 & 64.8 $\pm$ 3.8 & 1086 & 63.8 $\pm$ 0.8 & 2773 \\
\hline
\multirow{3}{*}{\makecell{Unanimous\\Consensus}}
& 3 & 60.6 $\pm$ 4.8 & 1670 & 37.2 $\pm$ 3.3 & 2336 & \textbf{68.0 $\pm$ 2.9} & 516 & 64.6 $\pm$ 3.6 & 1760 \\
& 5 & 61.6 $\pm$ 1.5 & 3573 & 40.2 $\pm$ 4.1 & 4362 & 66.4 $\pm$ 4.2 & 1190 & 64.0 $\pm$ 1.2 & 3631 \\
& 7 & 58.6 $\pm$ 4.3 & 5768 & 43.0 $\pm$ 1.6 & 6727 & 64.4 $\pm$ 3.5 & 2125 & 63.8 $\pm$ 1.3 & 5473 \\
\hline
\multirow{3}{*}{\makecell{Stance\\Entropy}}
& 3 & 62.0 $\pm$ 3.4 & 1287 & 37.8 $\pm$ 3.9 & 1516 & 67.6 $\pm$ 2.7 & 464 & 63.8 $\pm$ 1.9 & 847 \\
& 5 & 59.4 $\pm$ 2.5 & 2258 & 42.4 $\pm$ 4.5 & 2126 & 66.8 $\pm$ 4.1 & 858 & 64.0 $\pm$ 1.6 & 1483 \\
& 7 & \textbf{63.2 $\pm$ 2.5} & 2918 & \textbf{45.0 $\pm$ 3.1} & 3150 & 64.4 $\pm$ 3.2 & 1148 & 63.0 $\pm$ 1.9 & 1928 \\
\hline
\multirow{3}{*}{\makecell{Verbal\\Confidence}}
& 3 & 60.6 $\pm$ 5.0 & 3595 & 35.0 $\pm$ 2.5 & 4112 & 65.0 $\pm$ 2.2 & 2809 & 64.0 $\pm$ 2.9 & 3995 \\
& 5 & 59.2 $\pm$ 1.5 & 6311 & 39.0 $\pm$ 7.0 & 6933 & 63.4 $\pm$ 4.7 & 4822 & 64.6 $\pm$ 3.4 & 6865 \\
& 7 & 62.6 $\pm$ 3.2 & 8992 & 42.4 $\pm$ 2.3 & 9953 & 63.2 $\pm$ 4.4 & 6988 & 62.4 $\pm$ 1.3 & 9583 \\
\hline
\multirow{3}{*}{\makecell{GPT-5-mini\\as Judge}}
 & 3 & 67.2 $\pm$ 1.9 & 1309 & 39.2 $\pm$ 2.8 & 1912 & \textbf{68.2$^{\star}$ $\pm$ 2.4} & 792 & 64.8 $\pm$ 1.9 & 1889 \\
 & 5 & \textbf{68.4$^{\star}$ $\pm$ 2.6} & 2509 & 44.0 $\pm$ 2.1 & 3269 & 67.2 $\pm$ 3.9 & 1547 & 64.2 $\pm$ 1.2 & 3456 \\
 & 7 & 68.2 $\pm$ 2.3 & 4023 & \textbf{45.2$^{\star}$ $\pm$ 3.4} & 4550 & 64.8 $\pm$ 3.1 & 2421 & 63.2 $\pm$ 1.0 & 4788 \\
\hline
\end{tabular}
\label{tab:phi4_all_terminations}
\end{table*}

\begin{table*}[htb!]
\caption{Llama-3.1: Debate accuracy (in \%) with standard deviation and average generated token count per sample across decision protocols.}
\centering
\setlength{\tabcolsep}{3pt} 
\begin{tabular}{l | c | cc | cc | cc | cc }
\hline
\multirow{2}{*}{Protocol} & \multirow{2}{*}{\#Ag.}
& \multicolumn{2}{c|}{MMLU}
& \multicolumn{2}{c|}{MMLU-Pro}
& \multicolumn{2}{c|}{StrategyQA}
& \multicolumn{2}{c}{CUAD} \\
& 
& Accuracy & Token
& Accuracy & Token
& Accuracy & Token
& Accuracy & Token \\
\hline
Single Agent & 1 & 69.4 $\pm$ 2.7 & 323 & 40.4 $\pm$ 2.3 & 458 & 67.0 $\pm$ 4.9 & 291 & \textbf{68.8 $\pm$ 4.4} & 272 \\
\hline
\multirow{3}{*}{\makecell{Early\\Majority}}
& 3 & 72.2 $\pm$ 1.3 & 710 & 45.0 $\pm$ 2.8 & 1206 & 67.0 $\pm$ 4.2 & 459 & 67.2 $\pm$ 2.8 & 637 \\
& 5 & \textbf{72.8 $\pm$ 3.2} & 1108 & \textbf{47.2 $\pm$ 1.3} & 1698 & 67.0 $\pm$ 2.3 & 767 & 64.6 $\pm$ 2.2 & 1065 \\
& 7 & 70.8 $\pm$ 4.0 & 1627 & 46.4 $\pm$ 2.9 & 2602 & 66.4 $\pm$ 1.1 & 1102 & 65.8 $\pm$ 3.6 & 1416 \\
\hline
\multirow{3}{*}{\makecell{Unanimous\\Consensus}}
& 3 & 71.2 $\pm$ 3.5 & 1178 & 45.2 $\pm$ 4.1 & 1707 & 63.2 $\pm$ 7.3 & 873 & 65.4 $\pm$ 2.1 & 1113 \\
& 5 & 72.4 $\pm$ 2.9 & 1840 & \textbf{47.2 $\pm$ 2.9} & 2580 & 66.2 $\pm$ 1.8 & 1527 & 65.4 $\pm$ 2.2 & 1617 \\
& 7 & 71.4 $\pm$ 3.9 & 2541 & 46.6 $\pm$ 3.5 & 4119 & 66.4 $\pm$ 2.4 & 2077 & 66.8 $\pm$ 3.1 & 2379 \\
\hline
\multirow{3}{*}{\makecell{Stance\\Entropy}}
& 3 & 71.4 $\pm$ 2.1 & 995 & 45.0 $\pm$ 1.6 & 940 & 64.0 $\pm$ 6.8 & 759 & 67.2 $\pm$ 2.3 & 897 \\
& 5 & 71.8 $\pm$ 3.1 & 1067 & 47.0 $\pm$ 1.9 & 1341 & 66.4 $\pm$ 2.6 & 1280 & 65.2 $\pm$ 2.9 & 1384 \\
& 7 & 71.0 $\pm$ 3.0 & 1559 & 47.0 $\pm$ 3.2 & 1860 & 66.6 $\pm$ 1.5 & 1366 & 66.6 $\pm$ 4.0 & 1665 \\
\hline
\multirow{3}{*}{\makecell{Verbal\\Confidence}}
& 3 & 70.0 $\pm$ 5.3 & 3040 & 44.4 $\pm$ 4.3 & 3218 & 63.2 $\pm$ 5.4 & 2255 & 64.2 $\pm$ 2.3 & 3318 \\
& 5 & 70.4 $\pm$ 1.7 & 4683 & 47.0 $\pm$ 5.2 & 5191 & 66.4 $\pm$ 3.2 & 3921 & 64.8 $\pm$ 2.2 & 5013 \\
& 7 & 68.8 $\pm$ 2.2 & 6728 & 46.4 $\pm$ 3.6 & 7295 & \textbf{67.4 $\pm$ 2.1} & 5884 & 66.4 $\pm$ 3.5 & 6918 \\
\hline

\multirow{3}{*}{\makecell{GPT-5-mini\\as Judge}}
 & 3 & 72.6 $\pm$ 1.2 & 1960 & 46.0 $\pm$ 1.9 & 2699 & \textbf{67.6$^{\star}$ $\pm$ 3.6} & 2052 & \textbf{69.8$^{\star}$ $\pm$ 3.3} & 1915 \\
 & 5 & \textbf{73.2$^{\star}$ $\pm$ 2.9} & 3029 & \textbf{48.8$^{\star}$ $\pm$ 2.5} & 4068 & 65.2 $\pm$ 2.3 & 3075 & 65.6 $\pm$ 2.0 & 2707 \\
 & 7 & 71.4 $\pm$ 3.4 & 3807 & 46.8 $\pm$ 3.1 & 5418 & 66.0 $\pm$ 1.8 & 3783 & 66.2 $\pm$ 2.7 & 3587 \\
\hline
\end{tabular}
\label{tab:llama-3.1_all_terminations}
\end{table*}

\begin{table*}[htb!]
\caption{Qwen-2.5: Debate accuracy (in \%) with standard deviation and average generated token count per sample across decision protocols.}
\centering
\setlength{\tabcolsep}{3pt} 
\begin{tabular}{l | c | cc | cc | cc | cc }
\hline
\multirow{2}{*}{Protocol} & \multirow{2}{*}{\#Ag.}
& \multicolumn{2}{c|}{MMLU}
& \multicolumn{2}{c|}{MMLU-Pro}
& \multicolumn{2}{c|}{StrategyQA}
& \multicolumn{2}{c}{CUAD} \\
& 
& Accuracy & Token
& Accuracy & Token
& Accuracy & Token
& Accuracy & Token \\
\hline
Single Agent & 1 & 71.6 $\pm$ 1.9 & 287 & 51.4 $\pm$ 2.3 & 443 & \textbf{71.6$^{\star}$ $\pm$ 1.7} & 219 & \textbf{72.8$^{\star}$ $\pm$ 1.9} & 302 \\
\hline
\multirow{3}{*}{\makecell{Early\\Majority}}
& 3 & 74.2 $\pm$ 1.1 & 411 & 53.8 $\pm$ 2.8 & 998 & 68.8 $\pm$ 4.9 & 347 & 71.0 $\pm$ 3.1 & 479 \\
& 5 & 76.2 $\pm$ 4.0 & 735 & 53.4 $\pm$ 4.0 & 1646 & 70.2 $\pm$ 2.8 & 599 & 69.4 $\pm$ 1.7 & 831 \\
& 7 & 75.8 $\pm$ 3.1 & 1083 & 52.6 $\pm$ 1.9 & 2176 & 69.0 $\pm$ 3.2 & 875 & 70.8 $\pm$ 3.0 & 1175 \\
\hline
\multirow{3}{*}{\makecell{Unanimous\\Consensus}}
& 3 & 74.2 $\pm$ 1.8 & 455 & 52.6 $\pm$ 3.0 & 1244 & 68.6 $\pm$ 6.1 & 389 & 70.6 $\pm$ 2.9 & 530 \\
& 5 & \textbf{77.4$^{\star}$ $\pm$ 3.6} & 844 & 53.0 $\pm$ 4.1 & 2045 & 69.8 $\pm$ 2.8 & 701 & 69.6 $\pm$ 1.7 & 989 \\
& 7 & 75.8 $\pm$ 3.3 & 1364 & 51.4 $\pm$ 2.2 & 2977 & 69.2 $\pm$ 4.5 & 1008 & 70.8 $\pm$ 3.0 & 1338 \\
\hline
\multirow{3}{*}{\makecell{Stance\\Entropy}}
& 3 & 74.0 $\pm$ 1.6 & 439 & 53.6 $\pm$ 2.6 & 854 & 68.6 $\pm$ 6.1 & 388 & 70.6 $\pm$ 2.9 & 512 \\
& 5 & 76.4 $\pm$ 4.0 & 717 & \textbf{54.0 $\pm$ 3.7} & 1343 & 69.8 $\pm$ 2.8 & 697 & 69.4 $\pm$ 1.7 & 936 \\
& 7 & 76.2 $\pm$ 3.1 & 1062 & 53.8 $\pm$ 1.3 & 1725 & 69.6 $\pm$ 4.2 & 904 & 70.8 $\pm$ 3.0 & 1210 \\
\hline
\multirow{3}{*}{\makecell{Verbal\\Confidence}}
& 3 & 74.0 $\pm$ 2.1 & 2395 & 53.0 $\pm$ 1.2 & 3330 & 70.2 $\pm$ 5.1 & 1949 & 70.2 $\pm$ 2.4 & 2733 \\
& 5 & 74.6 $\pm$ 3.3 & 4416 & \textbf{54.0 $\pm$ 2.6} & 5667 & 71.4 $\pm$ 2.8 & 3199 & 69.8 $\pm$ 1.5 & 4757 \\
& 7 & 73.6 $\pm$ 2.9 & 6510 & 53.0 $\pm$ 2.2 & 8037 & 71.4 $\pm$ 4.6 & 4544 & 72.4 $\pm$ 2.1 & 6707 \\
\hline

\multirow{3}{*}{\makecell{GPT-5-mini\\as Judge}}
 & 3 & 74.6 $\pm$ 1.2 & 636 & 54.6 $\pm$ 2.2 & 1268 & 68.6 $\pm$ 4.6 & 666 & 70.8 $\pm$ 2.6 & 650 \\
 & 5 & \textbf{77.4$^{\star}$ $\pm$ 4.0} & 1043 & \textbf{55.2$^{\star}$ $\pm$ 3.1} & 1848 & 70.0 $\pm$ 2.4 & 1106 & 69.4 $\pm$ 1.5 & 1184 \\
 & 7 & 77.2 $\pm$ 3.0 & 1571 & 54.2 $\pm$ 1.3 & 2567 & 69.6 $\pm$ 3.9 & 1426 & 70.8 $\pm$ 2.7 & 1526 \\
\hline

\end{tabular}
\label{tab:qwen-2.5_all_terminations}
\end{table*}

\begin{table*}[htb!]
\caption{GPT-5-mini: Debate accuracy (in \%) with standard deviation and average generated token count per sample across decision protocols.}
\centering
\setlength{\tabcolsep}{3pt} 
\begin{tabular}{l | c | cc | cc | cc | cc }
\hline
\multirow{2}{*}{Protocol} & \multirow{2}{*}{\#Ag.}
& \multicolumn{2}{c|}{MMLU}
& \multicolumn{2}{c|}{MMLU-Pro}
& \multicolumn{2}{c|}{StrategyQA}
& \multicolumn{2}{c}{CUAD} \\
& 
& Accuracy & Token
& Accuracy & Token
& Accuracy & Token
& Accuracy & Token \\
\hline
Single Agent & 1 & 90.8 $\pm$ 0.4 & 550 & 82.8 $\pm$ 1.8 & 979 & 77.0 $\pm$ 1.6 & 809 & \textbf{83.4$^{\star}$ $\pm$ 1.5} & 670 \\
\hline
\multirow{3}{*}{\makecell{Early\\Majority}}
& 3 & 91.2 $\pm$ 0.8 & 1369 & 82.6 $\pm$ 1.3 & 2292 & 77.0 $\pm$ 1.2 & 1630 & 80.6 $\pm$ 0.5 & 1719 \\
& 5 & 90.8 $\pm$ 0.8 & 2255 & 83.2 $\pm$ 2.0 & 3592 & 77.2 $\pm$ 2.2 & 2632 & 79.8 $\pm$ 1.5 & 2769 \\
& 7 & \textbf{92.6$^{\star}$ $\pm$ 1.5} & 3126 & 82.6 $\pm$ 1.3 & 4656 & 75.4 $\pm$ 1.3 & 3651 & 80.2 $\pm$ 1.3 & 3881 \\
\hline
\multirow{3}{*}{\makecell{Unanimous\\Consensus}}
& 3 & 91.0 $\pm$ 1.2 & 1388 & 82.8 $\pm$ 1.3 & 2457 & 76.8 $\pm$ 1.1 & 1671 & 80.4 $\pm$ 0.5 & 1804 \\
& 5 & 91.4 $\pm$ 1.1 & 2293 & 83.2 $\pm$ 1.8 & 3659 & 77.0 $\pm$ 2.1 & 2778 & 80.0 $\pm$ 1.9 & 2934 \\
& 7 & 92.4 $\pm$ 1.7 & 3214 & 82.8 $\pm$ 1.3 & 4989 & 75.0 $\pm$ 1.0 & 3855 & 80.2 $\pm$ 1.5 & 4183 \\
\hline
\multirow{3}{*}{\makecell{Stance\\Entropy}}
& 3 & 91.0 $\pm$ 1.2 & 1388 & 82.6 $\pm$ 1.3 & 2292 & 76.8 $\pm$ 1.1 & 1665 & 80.4 $\pm$ 0.5 & 1796 \\
& 5 & 91.0 $\pm$ 0.7 & 2261 & 83.2 $\pm$ 2.0 & 3565 & 77.0 $\pm$ 2.1 & 2760 & 80.0 $\pm$ 1.9 & 2934 \\
& 7 & \textbf{92.6$^{\star}$ $\pm$ 1.5} & 3132 & 82.6 $\pm$ 1.3 & 4656 & 75.0 $\pm$ 1.0 & 3714 & 80.2 $\pm$ 1.5 & 4034 \\
\hline
\multirow{3}{*}{\makecell{Verbal\\Confidence}}
& 3 & 91.0 $\pm$ 1.2 & 5885 & 82.6 $\pm$ 0.5 & 8521 & 77.2 $\pm$ 1.3 & 6937 & 80.4 $\pm$ 1.3 & 7473 \\
& 5 & 91.2 $\pm$ 1.5 & 9927 & \textbf{83.4 $\pm$ 1.7} & 13619 & \textbf{77.8$^{\star}$ $\pm$ 1.3} & 11661 & 80.4 $\pm$ 1.5 & 12199 \\
& 7 & 92.0 $\pm$ 1.0 & 13947 & 82.8 $\pm$ 1.3 & 18425 & 75.4 $\pm$ 1.7 & 16349 & 81.8 $\pm$ 2.2 & 16885 \\
\hline

\multirow{3}{*}{\makecell{GPT-5-mini\\as Judge}}
 & 3 & 91.2 $\pm$ 0.7 & 1423 & 83.2 $\pm$ 1.2 & 2435 & 76.8 $\pm$ 1.0 & 1721 & 80.2 $\pm$ 0.7 & 1876 \\
 & 5 & 91.0 $\pm$ 1.1 & 2561 & \textbf{84.0$^{\star}$ $\pm$ 2.1} & 3652 & 77.0 $\pm$ 1.9 & 2873 & 79.8 $\pm$ 1.3 & 2988 \\
 & 7 & 92.4 $\pm$ 1.5 & 3290 & 83.0 $\pm$ 1.1 & 4849 & 75.0 $\pm$ 0.9 & 3970 & 80.4 $\pm$ 1.5 & 4218 \\
\hline
\end{tabular}
\label{tab:gpt-5-mini_all_terminations}
\end{table*}

\begin{table*}[htb!]
\caption{Heterogeneous agents with Phi-4, Llama-3.1, Qwen-2.5: Debate accuracy (in \%) with standard deviation and average generated token count per sample across decision protocols.}
\centering
\setlength{\tabcolsep}{3pt} 
\begin{tabular}{l | c | cc | cc | cc | cc }
\hline
\multirow{2}{*}{Protocol} & \multirow{2}{*}{\#Ag.}
& \multicolumn{2}{c|}{MMLU}
& \multicolumn{2}{c|}{MMLU-Pro}
& \multicolumn{2}{c|}{StrategyQA}
& \multicolumn{2}{c}{CUAD} \\
& 
& Accuracy & Token
& Accuracy & Token
& Accuracy & Token
& Accuracy & Token \\
\hline
\multirow{3}{*}{\makecell{Early\\Majority}}
& 3 & 73.0 $\pm$ 2.2 & 728 & 50.2 $\pm$ 3.8 & 1585 & 70.2 $\pm$ 1.3 & 436 & 69.8 $\pm$ 2.0 & 744 \\
& 5 & 76.0 $\pm$ 1.4 & 1170 & \textbf{53.8 $\pm$ 3.0} & 1979 & 68.2 $\pm$ 1.3 & 792 & 68.6 $\pm$ 2.9 & 1134 \\
& 7 & 75.8 $\pm$ 1.3 & 1720 & 50.6 $\pm$ 3.5 & 2919 & 69.2 $\pm$ 1.3 & 1128 & 69.8 $\pm$ 1.3 & 1642 \\
\hline
\multirow{3}{*}{\makecell{Unanimous\\Consensus}}
& 3 & 73.4 $\pm$ 1.3 & 1097 & 48.0 $\pm$ 4.9 & 2536 & 69.6 $\pm$ 1.8 & 609 & 69.6 $\pm$ 2.5 & 1193 \\
& 5 & \textbf{76.6 $\pm$ 2.4} & 1882 & 53.6 $\pm$ 2.9 & 4030 & 67.8 $\pm$ 2.9 & 1283 & 70.2 $\pm$ 3.6 & 1951 \\
& 7 & 76.2 $\pm$ 1.3 & 3406 & 50.8 $\pm$ 4.0 & 6256 & 70.2 $\pm$ 0.8 & 1860 & 70.2 $\pm$ 1.1 & 3481 \\
\hline
\multirow{3}{*}{\makecell{Stance\\Entropy}}
& 3 & 73.6 $\pm$ 1.8 & 798 & 51.0 $\pm$ 4.2 & 1012 & 69.2 $\pm$ 0.8 & 535 & 70.8 $\pm$ 1.5 & 860 \\
& 5 & 76.2 $\pm$ 1.8 & 1093 & 53.6 $\pm$ 3.6 & 1432 & 68.0 $\pm$ 2.9 & 1102 & 69.8 $\pm$ 3.1 & 1446 \\
& 7 & 75.6 $\pm$ 1.8 & 1614 & 51.0 $\pm$ 3.5 & 1924 & 70.0 $\pm$ 1.4 & 1286 & 70.0 $\pm$ 1.4 & 1866 \\
\hline
\multirow{3}{*}{\makecell{Verbal\\Confidence}}
& 3 & 73.8 $\pm$ 2.4 & 3526 & 51.6 $\pm$ 5.2 & 4169 & \textbf{71.0$^{\star}$ $\pm$ 2.7} & 2474 & 69.8 $\pm$ 1.3 & 3870 \\
& 5 & 72.8 $\pm$ 3.6 & 5936 & 51.2 $\pm$ 3.0 & 6595 & 70.2 $\pm$ 2.8 & 4557 & \textbf{71.6$^{\star}$ $\pm$ 3.2} & 6149 \\
& 7 & 73.8 $\pm$ 1.8 & 8640 & 50.0 $\pm$ 2.9 & 9406 & 68.4 $\pm$ 2.3 & 6489 & 69.0 $\pm$ 2.7 & 8754 \\
\hline
\multirow{3}{*}{\makecell{GPT-5-mini\\as Judge}}
 & 3 & 73.0 $\pm$ 1.4 & 1614 & 49.8 $\pm$ 5.4 & 3134 & 69.4 $\pm$ 1.0 & 1265 & 70.6 $\pm$ 1.5 & 1661 \\
 & 5 & \textbf{77.0$^{\star}$ $\pm$ 2.4} & 2547 & \textbf{54.0$^{\star}$ $\pm$ 2.4} & 4470 & 68.8 $\pm$ 0.4 & 2248 & 69.8 $\pm$ 2.8 & 2618 \\
 & 7 & 75.6 $\pm$ 1.9 & 3695 & 49.4 $\pm$ 3.7 & 6237 & 69.2 $\pm$ 1.8 & 3001 & 69.8 $\pm$ 1.6 & 3931 \\
\hline
\end{tabular}
\label{tab:heterogeneous_all_terminations}
\end{table*}

\end{document}